\documentclass[a4paper,11pt,dvipsnames]{article}
\usepackage{pos}

\usepackage{xcolor}
\usepackage{tikz}
\usepackage{caption}
\usepackage{tensor}

\title{Doubly Special Relativity and Relative Locality}

\author[b]{Andrea Bevilacqua}
\author*[a,b]{Jerzy Kowalski-Glikman}

\affiliation[a]{University of Wroclaw, Faculty of Physics and Astronomy,\\
Pl.\ Maksa Borna 9, Wroclaw, Poland}

\affiliation[b]{National Centre for Nuclear Research,\\
  Pateura 7, Warsaw, Poland}

\emailAdd{andrea.bevilacqua@ncbj.gov.pl}
\emailAdd{jerzy.kowalski-glikman@uwr.edu.pl}

\abstract{This is a write-up of two lectures delivered at COST CA18108 First Training School. They cover the motivations and some basic technical results in the field of Doubly Special Relativity and Relative Locality. The energy-dependent speed of light is a recurring theme here. The soccer ball problem is also briefly described. The video recording of these and other lectures can be found at \url{https://www.youtube.com/channel/UChRfzgjTkO9sOkwq358kHZw/videos}. }

\FullConference{%
COST CA18108 First Training School\\
27 September--5 October, 2021\\
Corfu, Greece
}

\begin{document}
\maketitle

\section{Motivations}

Without doubts the construction of a quantum theory of gravity is the single most pressing challenge the contemporary high energy physics community is facing.  We still do not know exactly what this theory is; however we can investigate some particular limits of it and build toy models, whose predictions might be, as I am going to argue here, even tested with current and near future observational technologies (see \cite{Addazi:2021xuf} for a recent comprehensive review.) Let us investigate how this limit, called the \textit{relative locality regime} comes about.

Quantum gravity is assumed to unify the two most successful theories of twentieth century physics,  General Relativity and Quantum Mechanics. These theories are characterized  by  dimensionful constants, $G_N$ and $\hbar$,  respectively, and describe systems and processes for which these constants are large, in an appropriate sense. For example, General Relativity describes systems for which the size $L$ and mass $M$ satisfy $R\sim G_N M$, and Quantum Mechanics is relevant when the action is of order $\hbar$. Similarly, Quantum Gravity describes systems/effect for which  $G_N$ and $\hbar$  are large at the same time, meaning, essentially, that  the Compton wavelength is of order of Schwarzschild radius. For the system of energy $E$ in the rest frame this means that $\hbar/E \sim G_N E$.

It is therefore commonly believed that a quantum theory of gravity, in which both GR and QM will play a fundamental role, becomes relevant when the scales of distance and energy are at the same time of the order of the Planck length and the Planck energy respectively
\begin{align}
	l_P = \sqrt{\hbar G_N} \approx 10^{-35} m
	\qquad
	E_P = \sqrt{\frac{\hbar}{G_N}} \approx 10^{19} GeV
\end{align}
In what follows, we will call $\kappa$ the energy scale of the order of $E_P$ at which quantum gravitational effects begin to become relevant.
For example, a scattering with impact parameter $l_P$ between two particles of energy $\kappa$ is expected to be best described by a quantum theory of gravity. Indeed one can see that when the impact parameter of the Planck energy particles scattering approaches $l_P$ then new phenomena begin to appear that requires full quantum gravity to understand (see e.g. \cite{Giddings:2011xs} and references therein.) Unfortunately, a description of such phenomena is still beyond reach.

One can however study the `relative locality regime' \cite{Amelino-Camelia:2011lvm}, characterized by the condition that the size of the Planck energy system is much larger than the Planck length
\begin{align}
	l \gg l_P 
	\qquad
	\text{\textit{and}}
	\qquad
	E \approx \kappa
\end{align}
This regime can be obtained as a limit of quantum gravity by sending both $\hbar$ and $G_N$ to zero in such a way that their ratio remains constant. In physical terms, it is expected to describe phenomena at energies comparable to Planck energies, with the characteristic size much larger than the Planck length, so that the spacetime foam effects could be safely ignored.

In the relative locality regime, by the principle of correspondence, we expect to have to do with a somehow modified but still relativistic theory. Indeed both in the low energy limit and in the ultraviolet one we have to do with the theories possessing relativistic symmetries (quantum field theory at the one end and quantum gravity at the another). This regime can be modelled by a theory that possesses two observer independent scales: one of velocity $c$\footnote{In what follows we will set $c=1$.}, since the theory is to be relativistic,  and one of mass/energy/momentum $\kappa$, which reflects the presence of the energy scale, a remnant of quantum gravity. Such theory was proposed two decades ago under the name of `Doubly (or Deformed) Special Relativity' \cite{Amelino-Camelia:2000cpa}, \cite{Amelino-Camelia:2000stu}, \cite{Kowalski-Glikman:2001vvk}. The introduction of an energy scale to relativistic theories is a nontrivial endeavour because the standard Lorentz transformations change energy. Therefore, in order to keep all the relativistic symmetries while at the same time incorporating an invariant energy scale, one needs to appropriately modify the Lorentz transformations as well as composition laws of various physical quantities. 

This line of reasoning is analogous to the historical development of relativity in physics. The principle of relativity was first formulated by Galileo, who stated that the laws of physics have the same form in any inertial reference frame. Since in Galilean relativity there is no observer-independent velocity scale, the only possible velocity composition law is linear (just for dimensional reasons, since only in the case of linear composition $V = v+u$ one does not need to introduce a velocity scale to make the dimensions of terms match). At the same time, the transformation laws between different inertial observer are also linear in the velocity. The principle of relativity was later extended by Einstein, who incorporated a fundamental scale of velocity $c$. Because of this, now a new composition of 3-velocities is possible, and in the case of Special Relativity the formula is highly non-linear, non-symmetric and non-associative
\begin{align}
	\mathbf{v}\oplus \mathbf{u}
	=
	\frac{1}{1 + \frac{\mathbf{u}\mathbf{v}}{c^2}}
	\left(
	\mathbf{v}
	+
	\frac{\mathbf{u}}{\gamma_{\mathbf{v}}}
	+
	\frac{1}{c^2}
	\frac{\gamma_{\mathbf{v}}}{1 + \gamma_{\mathbf{v}}}
	(\mathbf{v}\mathbf{u})\mathbf{v}
	\right)
	\qquad
	\qquad
	\gamma_{\mathbf{v}}
	=
	\sqrt{1 - \frac{\mathbf{v}^2}{c^2}}
\end{align}
At the same time, in order to preserve the relativity principle, Galilean transformations were extended into Lorentz transformations. The addition of a new invariant mass scale $\kappa$ gives  rise to a yet new framework called Doubly Special Relativity (DSR in short), and once again this gives rise, this time, to modified energy and momentum composition laws as well as deformed Poincar\'e symmetry, which will be described in these lecture notes. As we will also see, the presence of the invariant scale $\kappa$ enforces the relaxation of the absolute locality postulate of Special Relativity, and locality becomes relative to the observer.

The aim is to study possible quantum-gravitational effects which become important at the relative locality regime. We will see that such deformed model predict results which slightly deviate from the ones obtained from Special Relativity. One such prediction is the different time of arrival of photons of different energy which have been simultaneously emitted by a distant source. Such a phenomenon can be matched against measurements in order to look for deviations from Special Relativity.

The reader interested in a more detailed description  of the topics briefly discussed in this notes may consult a recent monograph \cite{Arzano:2021scz}.

\section{Relativistic point particle and LIV models}

We start from the standard action for a point-like relativistic particle, written in the first order formalism (linear in velocities) 
\begin{align}\label{action}
	S = \int d\tau \dot{x}^\mu p_\mu - N(p^2 + m^2)
\end{align}
where we are using the signature $-+++$. The equations of motion (EoM) resulting from variations over $x$, $p$, and $N$ are given by
\begin{align}\label{undefaction}
	\dot{p}_\mu = 0
	\qquad
	\dot{x}^\mu = 2Np^\mu
	\qquad
	p^2 = -E^2 + \mathbf{p}^2 = -m^2.
\end{align}
The variable $N$ is a kind of gauge variable and can be fixed a posteriori. As an example, for a massive particle it can be fixed to $N=\frac{1}{2m}$ so that the EoM for $x^\mu$ become 
$\dot{x}^\mu = \frac{p^\mu}{m}$. One can easily compute the speed of light in this model because we now have $\dot{x}^\mu \propto p^\mu$, and for a massless particle $p^2 = 0$, hence
\begin{align}
	v^2 = 
	\left(
	\frac{d \mathbf{x}}{d x^0}
	\right)^2
	=
	\left(
	\frac{\dot{\mathbf{x}}}{\dot{x}^0}
	\right)^2
	=
	\left(
	\frac{\mathbf{p}^2}{p_0^2}
	\right)_{\mathbf{p}^2 = p_0^2}
	=
	1
\end{align}
The symmetries that leave the the action \eqref{action} invariant and the EoM \eqref{undefaction} covariant are\footnote{We encourage the reader to check it explicitly.}  translations
\begin{align}
    \delta x^\mu = \epsilon^\mu\,,
\end{align}
rotations 
\begin{align}
	\delta x^i = \rho^k \tensor{\epsilon}{^i_{jk}} x^j
	\qquad
	\delta x^0 = 0
	\qquad
	\delta p_i = \rho^k \tensor{\epsilon}{_i^j_k} p_j
	\qquad
	\delta p_0 = 0\,,
\end{align}
and boosts
\begin{align}
	\delta x^i = -\lambda^i x^0
	\qquad
	\delta x^0 = -\lambda_i x^i
	\qquad
	\delta p_i = \lambda_i p_0
	\qquad
	\delta p_0 = \lambda^i p_i\,.
\end{align}

We have now 10 independent infinitesimal transformations associate with 10 independent parameters $\{\epsilon^\mu, \rho^k, \lambda^i\}$, and therefore a 10-dimensional algebra of the symmetry group. We can rewrite the above transformations in a more abstract way, separating the generators (which contain the actual physical information about the transformations) from the parameters in the following way
\begin{align}\label{abstractvariations}
	\delta_T(\bullet) 
	&= 
	\epsilon^\mu P_\mu \triangleright \bullet  
	\qquad \qquad \qquad \qquad 
	P_\mu \triangleright x^\nu = \delta_\mu^\nu
	\\
	\delta_R(\bullet) 
	&= 
	\rho^i R_i \triangleright \bullet 
	\qquad \text{acting as} \qquad \, \, \,
	R_i \triangleright x^j = \tensor{\epsilon}{^j_{ik}} x^k\\
	\delta_N(\bullet) 
	&= 
	\lambda^i N_i \triangleright \bullet
	\qquad \qquad \qquad \qquad \, \,
	N_i \triangleright x^j = \delta_i^j x^0
\end{align}
where the generators $\{P_\mu, R_i, N_i\}$ satisfy the Poincar\'e algebra 
\begin{align}
	[P_\mu, P_\nu] = 0
	\qquad
	[R_i, P_j] = i \tensor{\epsilon}{_{ij}^k}P_k
	\qquad
	[R_i, P_0] = 0
	\qquad
	[N_i, P_j] = -i \eta_{ij} P_0 \nonumber 
\end{align}
\begin{align}
	[N_i, P_0] = -i P_i
	\qquad
	[R_i, R_j] = i \tensor{\epsilon}{_{ij}^k} R_k 
	\qquad
	[R_i, N_j] = i \tensor{\epsilon}{_{ij}^k} N_k 
	\qquad
	[N_i, N_j] = -i \tensor{\epsilon}{_{ij}^k} R_k \nonumber 
\end{align}
So far we discussed the relativistic particle theory that preserved all the relativistic symmetries of Poincar\'e algebra/group. The simplest way to obtain a Lorentz-invariance-violating (LIV) theory is to keep the kinetic term in the action  \eqref{undefaction} as it is, changing the dispersion relation to the modified form
\begin{align*}
	p^2 + m^2 \, \rightarrow \, \mathcal{C}_\kappa(p) + m^2
\end{align*}
where $\mathcal{C}_\kappa(p)$ must satisfy the following conditions
\begin{itemize}
	\item It cannot depend only on $p^2$ (otherwise the relativistic symmetries will be preserved, and, in particular  the speed of massless particles will still be equal 1 -- the reader is encouraged to check this statement);
	\item $	\lim_{\kappa \rightarrow \infty} \mathcal{C}_\kappa (p) = p^2$, which is just the natural complementarity condition.
\end{itemize}
The modified mass-shell condition $\mathcal{C}_\kappa (p)=-m^2$  is therefore assumed to be not Lorentz invariant anymore. Using the momentum representation of the boost, we can in general write this as a requirement
\begin{align}
	p_0 \frac{\partial \mathcal{C}_\kappa(p)}{\partial \mathbf{p}^i} + \mathbf{p}_i \frac{\partial \mathcal{C}_\kappa(p)}{\partial p_0} \neq 0
\end{align}
How could this kind of model arise? An idea is that the quantum gravity vacuum (for example string theory vacuum), describing the spacetime in which we live in, is assumed to violate Lorentz symmetry \cite{Kostelecky:1988zi}, \cite{Colladay:1998fq}, \cite{Mattingly:2005re}. This Lorentz invariance violation must disappear at low energies, which is the reason for the conditions that in the limit $\kappa\rightarrow \infty$ the mass shell condition returns to its special-relativistic form.

The equations of motion  for photons in such a model are given by
\begin{align*}
	\dot{p}_\mu = 0
	\qquad
	\dot{x}^\mu = N \frac{\partial \mathcal{C}_\kappa}{\partial p_\mu}
	\qquad
	\mathcal{C}_\kappa (p) = 0
\end{align*}
which imply that in general
\begin{align}
	v^2 = 
	\left(
	\frac{\dot{\mathbf{x}}}{\dot{x}^0}
	\right)^2
	=
	\left(
	\frac{\partial \mathcal{C}_\kappa(p)}{\partial \mathbf{p}_i}
	\Big/
	\frac{\partial \mathcal{C}_\kappa(p)}{\partial p_0}
	\right)_{\mathcal{C}_\kappa = 0}
	\neq 1
\end{align}
and $v$ is momentum dependent. In physical terms this means that photons of different energies move with different velocities. This effect is very small, dumped by some power of the ratio $E/\kappa$, but in principle measurable if the source is at cosmological distances \cite{Amelino-Camelia:1997ieq}.

\section{From deformed dispersion relations to curved momentum space}

After the previous discussion about LIV models it is natural to ask ourselves the following question: is it possible to choose $\mathcal{C}_\kappa(p)$ in such a way that the Lorentz transformations are \textit{not} violated anymore (in the appropriate sense)?
The answer is surprisingly affirmative for a very large class of possible $\mathcal{C}_\kappa(p)$. The reasoning goes as follows. Let us assume for simplicity that $\mathcal{C}_\kappa(p)$ is invariant under rotations\footnote{This is just a simplifying assumption and not a fundamental requirement for the validity of the following reasoning.}, so that is is function of $p_0, \mathbf{p}^2$. Then, consider the generalized boost
\begin{align}\label{defboost}
	N_i^\kappa
	=
	A(p_0, \mathbf{p}^2) \frac{\partial}{\partial \mathbf{p}_i}
	+
	B_i(p_0, \mathbf{p}^2) \frac{\partial}{\partial p_0}
\end{align}
and impose that it satisfy the requirements
\begin{align}\label{requirements}
	[N_i^\kappa, N_j^\kappa] \propto \tensor{\epsilon}{_{ij}^k}R_k
	\qquad 
	[R_i^\kappa, N_j^\kappa] \propto \tensor{\epsilon}{_{ij}^k}N_k
	\qquad
	N_i^\kappa \triangleright \mathcal{C}_\kappa(p_0, \mathbf{p}^2)=0
\end{align}
It turns out that the requirements in  \eqref{requirements} can be easily satisfied \cite{Kowalski-Glikman:2002eyl} so that the Lorentz symmetry algebra is \textit{not} modified and the dispersion relation is still relativistic invariant. It is important to notice, however that the kinetic term $\dot x^\mu p_\mu$ is \textit{not} invariant in general under such deformed transformations, and it should be properly modified to render the whole action invariant. 

This surprising fact can be understood if one realizes that  the deformation of the boost and of the dispersion relation can be associated with coordinate changes in the momentum space, which, in the presence of the scale $\kappa$, are not necessarily confined to the linear ones. A historically relevant example (which we will also use later) of such a coordinate transformation is give by $p \mapsto p(k)$
\begin{align}\label{changeofcoord}
	p_0(k_0, \mathbf{k}) &= \kappa \sinh \frac{k_0}{\kappa}
	+
	\frac{\mathbf{k}^2}{2\kappa} e^{\frac{k_0}{\kappa}} \\
	\mathbf{p}(k_0, \mathbf{k})
	&=
	\mathbf{k} e^{\frac{k_0}{\kappa}}
\end{align}
where the deformed $\mathcal{C}_\kappa(p)$ is given by 
\begin{align}\label{defdisprel}
	\mathcal{C}_\kappa(p) = -2\kappa^2
	\sqrt{1+\frac{-p_0^2 + \mathbf{p}^2 }{\kappa^2} } +2 \kappa^2\simeq p_0^2 + O\left(\frac1{\kappa^2}\right)
\end{align}
which can be written in terms of the new coordinates $(k_0, \mathbf{k})$ as
\begin{align}
	\mathcal{C}_\kappa(k) = -4 \kappa^2 \sinh^2 
	\left(
	\frac{k_0}{2\kappa}
	\right)
	+
	\mathbf{k}^2 e^{\frac{k_0}{\kappa}}.
\end{align}
$\mathcal{C}_\kappa(k)$ is invariant under deformed boosts of the form of \eqref{defboost}, because $\mathcal{C}_\kappa(p)$ is invariant under standard boosts.
At this point, two important observations need to be made. 

The first  is that changes of coordinates cannot be physical. The classical argument involves a mad saboteur which is able to hack their way into the CERN systems with the aim of replacing each and every measured momentum with a function of the momenta themselves. Of course, the physics measured by the LHC experiment remains unchanged, although it is expressed in a different way.
Therefore deformation alone cannot bear any physical meaning, unless it is achieved in a way that it is not cancellable through a coordinate transformations. Furthermore, physical observables have to be momentum space coordinate independent objects.

The second observation is that, in the same way as one goes from special to general relativity by introducing a velocity scale and a nontrivial geometry of the (three) velocities manifold, one can very well conceive to go from a flat to a curved momentum space, as was first considered by Max Born around 1938. It is natural to look for  deformations of the dispersion relation and of the Lorentz transformations coming from the curvature of momentum space. In this way one could get a deformation inspired by the one described in  \eqref{defboost}, \eqref{changeofcoord}, \eqref{defdisprel}, which cannot be removed  by a coordinate transformation (since curvature is not coordinate dependent). The curvature of momentum space can naturally be associated to the scale $\kappa$ at hand. The idea to associate a fundamental scale to nontrivial geometry is also not new, and it was expressed by  Carl Friedrich  Gauss at the very dawn of differential geometry:
\begin{quote}
    The assumption that the sum of the three angles [of a triangle] is smaller than 180◦ leads to a geometry which is quite different from our (Euclidean) geometry, but which is in itself completely consistent. I have satisfactorily constructed this geometry for myself [\ldots ], except for the determination of one constant, which cannot be ascertained a priori. [\ldots] Hence I have sometimes in jest expressed the wish that Euclidean geometry is not true. For then we would have an absolute a priori unit of measurement\footnote{As cited in J. Milnor “Hyperbolic geometry: the first 150 years,”Bull. Am. Math. Soc.69 (1982).}.
\end{quote}
This idea can be summarized by saying that one can expect a non-trivial geometry if there is a scale which allows it. In other words, everything is curved unless it cannot be.

Motivated by  these considerations, we start now the more detailed studies of curved momentum space. The conditions that such a space must satisfy are simple: 
\begin{itemize}
\item[1)] The (deformed) Lorentz group must act on it;
\item[2)] The scale $\kappa$ must be related to to its geometry, for example being the curvature of momentum space;
\item[3)] In the limit $\kappa \rightarrow \infty$ one should get back the canonical, flat momentum space.
\end{itemize}
The simplest case that we can consider is the one of constant positive curvature $\frac{1}{\kappa^2}$, i.e. (some submanifold of) the de Sitter space\footnote{One could very well consider the interesting case of a momentum space whose curvature depends on the momentum, i.e. with non-trivial local curvature. However, in this case the scale $\kappa$ would not have an immediate fundamental meaning, and the treatment of such a case would be more difficult than the simpler case of constant curvature. }. But how to build such a momentum space? 

The idea is simple, and requires us to go back to eq. \eqref{changeofcoord}. We can use the coordinates $(p_0, \mathbf{p})$ which form a Lorentz vector, and add a fifth component $p_4$ which is a Lorentz scalar. In this five dimensional space, we then find out what is the manifold individuated by the coordinates $(k_0, \mathbf{k})$ defined in \eqref{changeofcoord}. Of course, also $p_4$ needs to be adequately expressed as a function of $k_0, \mathbf{k}$ in order to obtain the correct submanifold. More precisely, the map $(k_0, \mathbf{k})\mapsto (p_0(k_0, \mathbf{k}), \mathbf{p}(k_0, \mathbf{k}), p_4(k_0, \mathbf{k}))$ defines a surface in the 5-dimensional space described by the coordinates $(p_0, \mathbf{p}, p_4)$. 

In this way we obtain a 4-dimensional submanifold of the 5-dimensional space described by $(p_0,\mathbf{p}, p_4)$. In particular, the following coordinates 
\begin{align}\label{5dcoord}
	p_0(k_0, \mathbf{k}) &= \kappa \sinh \frac{k_0}{\kappa}
	+
	\frac{\mathbf{k}^2}{2\kappa} e^{\frac{k_0}{\kappa}} \nonumber  \\
	\mathbf{p}_i(k_0, \mathbf{k})
	&=
	\mathbf{k}_i e^{\frac{k_0}{\kappa}} \\
	p_4(k_0, \mathbf{k})
	&=
	\kappa \cosh \frac{k_0}{\kappa}
	-
	\frac{\mathbf{k}^2}{2\kappa} e^{\frac{k_0}{\kappa}} \nonumber 
\end{align}
satisfy
\begin{align}\label{deSitterrel}
	-p_0^2 + \mathbf{p}^2 + p_4^2 = \kappa^2
\end{align}
so that they describe a 4-dimensional submanifold of the de Sitter space \eqref{deSitterrel} defined by the condition
\begin{align}\label{deSitterre2}
	p_0+p_4 = \kappa e^{\frac{k_0}{\kappa}} >0.
\end{align}
Of course, $(k_0, \mathbf{k})$ are coordinates on this submanifold. In other words, the coordinates  $(k_0, \mathbf{k})$ cover half of de Sitter space. Incidentally, these coordinates are in the direct analogy with to the Friedman-Robertson-Walker-de Sitter universe. The reader is encouraged to check this explicitly by computing the induced metric on the submanifold \eqref{deSitterrel}, \eqref{deSitterre2} using \eqref{5dcoord} (or look at \eqref{momspacemetric} for the answer).

\section{Deformed action and its properties}

We now have a curved momentum space, and we want to write an action for a particle with momenta in this space.
To illustrate the process, we first generalize the action in eq. \eqref{action} to the case of curved spacetime. After this we will use a similar procedure to instead generalize \eqref{action} to the case of a curved momentum space.

The generalization of \eqref{action} to a curved momentum space is straightforward, and makes use of the vierbein (also called tetrad). In this section, we will use Latin letters $a, b, \dots$ to indicate flat spacetime (momentum space) indices, and Greek letters $\mu, \nu, \dots$ for curved spacetime (momentum space)\footnote{Using this convention, all the indices in eq. \eqref{action} have to be Latin indices.}. The vierbein is defined through the relation
\begin{align}
	g_{\mu\nu}(x) = \eta_{ab}\tensor{e}{_\mu^a}(x)\tensor{e}{_\nu^b}(x)
\end{align}
and using it we can write the action as follows
\begin{align}\label{actioncurvedst}
	S = \int d\tau \,\, \dot{x}^\mu \tensor{e}{_\mu^a}(x) p_a - N(\eta^{ab} p_a p_b + m^2)
\end{align}
which is the correct action for a pointlike particle in curved spacetime (but flat momentum space, as one can see from the dispersion relation). The equations of motion of this action give the standard geodesic equations in the metric $g_{\mu\nu}(x)$. This is easy to see by noticing that $p_a = \tensor{e}{^\mu_a}(x) p_\mu$ so that the action in \eqref{actioncurvedst} reduces to the more familiar
\begin{align}
	S = \int d\tau \,\, \dot{x}^\mu p_\mu - N(g^{\mu\nu}(x) p_\mu p_\nu + m^2)
\end{align}
whose equations of motion are now (ignoring for the moment the on-shell relation $g^{\mu\nu}(x) p_\mu p_\nu + m^2 = 0$)
\begin{align}
	\dot{x}^\mu =& 2N g^{\mu\nu}(x)p_\nu \\
	\dot{p}_\alpha + N (&\partial_\alpha g^{\mu\nu}) p_\mu p_\nu = 0.
\end{align}
Substituting the first one into the second we get 
\begin{align}
	\frac{d}{d\tau} 
	\left(
	\frac{1}{2N}
	g_{\alpha\mu} \dot{x}^\mu
	\right)
	+
	N
	\partial_\alpha g^{\mu\nu}
	\left(
	\frac{1}{2N}
	g_{\mu\beta} \dot{x}^\beta
	\right)
	\left(
	\frac{1}{2N}
	g_{\nu\gamma} \dot{x}^\gamma
	\right) = 0
\end{align}
which reduces to 
\begin{align}
	\frac{d}{d\tau} 
	\left(
	g_{\alpha\mu} \dot{x}^\mu
	\right)
	+
	\frac{1}{2}
	(\partial_\alpha g^{\mu\nu})
	g_{\mu\beta} \dot{x}^\beta
	g_{\nu\gamma} \dot{x}^\gamma
	= 0
\end{align}
and hence 
\begin{align}
	g_{\alpha\mu} \ddot{x}^\mu
	+ 
	\frac{1}{2}
	(\partial_\sigma g_{\alpha\mu} +
	\partial_\mu g_{\alpha\sigma} - \partial_\alpha g_{\mu\sigma})\dot{x}^\mu \dot{x}^\sigma
	=0
\end{align}
which are indeed the canonical geodesic equations.  Notice that we used the fact that 
\begin{align}
	(\partial_\alpha g^{\mu\nu})
	g_{\mu\beta}g_{\nu\gamma}
	=
	-\partial_\alpha g_{\beta\gamma}.
\end{align}
Coming back to our discussion, we now have an action in \eqref{actioncurvedst} that correctly represents the motion of a particle in curved spacetime and flat momentum space. It is now immediate to see how the same construct can be applied to the case of a flat spacetime but curved momentum space. We just need to introduce a non-trivial vierbein which allow us to proceed in the same manner. Therefore, we write the action
\begin{align}\label{actioncurvedms}
	S_\kappa = 
	-\int d\tau \,\, x^a \tensor{E}{_a^\mu}(k, \kappa) \dot{k}_\mu - N(\mathcal{C}_\kappa(k) + m^2).
\end{align}
Notice that now the time derivative, denoted by dot acts on the momentum and not the coordinate. In the flat space the placement of the dot is irrelevant if one ignores boundary contributions (since the choices differ just by a total derivative), but it is essential for what follows and relative locality to be discussed below that it acts on the momentum.  Notice that this action is written in terms of the coordinates $(k_0, \mathbf{k})$ which cover our curved momentum space. In order to obtain the  expression for $\tensor{E}{_a^\mu}(k, \kappa)$ we need the form of the metric $G$ in momentum space, and this is easily obtained from  \eqref{5dcoord} and \eqref{deSitterrel}. The five-dimensional line element is given by $ds_5^2 = -dp_0^2 + d\mathbf{p}^2 + dp_4^2$, and from this we can get the line element in the submanifold of the de Sitter surface simply by substituting \eqref{5dcoord}. We have
\begin{align}
		dp_0(k_0, \mathbf{k}) &= d k_0 \cosh \frac{k_0}{\kappa}
		+
		d\mathbf{k}\frac{\mathbf{k}}{\kappa}
		e^{\frac{k_0}{\kappa}}
		+
		\frac{\mathbf{k}^2}{2\kappa^2} e^{\frac{k_0}{\kappa}} dk_0
		\nonumber  \\
		d\mathbf{p}(k_0, \mathbf{k})
		&=
		d\mathbf{k} e^{\frac{k_0}{\kappa}} 
		+
		\frac{\mathbf{k}}{\kappa} e^{\frac{k_0}{\kappa}} dk_0
		\\
		dp_4(k_0, \mathbf{k})
		&=
		d k_0 \sinh \frac{k_0}{\kappa}
		-
		d\mathbf{k}\frac{\mathbf{k}}{\kappa}
		e^{\frac{k_0}{\kappa}}
		-
		\frac{\mathbf{k}^2}{2\kappa^2} e^{\frac{k_0}{\kappa}} dk_0 \nonumber 
\end{align}
and therefore
\begin{align}\label{momspacemetric}
	-dp_0^2 + d\mathbf{p}^2 + dp_4^2
	\mapsto
	dk_0^2
	\left(
	-\cosh^2 \frac{k_0}{\kappa}
	+
	\sinh^2 \frac{k_0}{\kappa}
	\right)
	+
	d\mathbf{k}^2 e^{2\frac{k_0}{\kappa}}
	=
	-d^2 k_0 
	+
	d\mathbf{k}^2 e^{2\frac{k_0}{\kappa}}
\end{align}
which means that $G = \text{diag} \left(-1,e^{2\frac{k_0}{\kappa}},e^{2\frac{k_0}{\kappa}},e^{2\frac{k_0}{\kappa}}\right)$ and, since the momentum space tetrad is defined in the usual way by\footnote{ Notice that since momenta are objects with lower indices, the metric in momentum space has upper indices.}
\begin{align}\label{momentumspacemetric}
	G^{\mu\nu}(k, \kappa) = \eta^{ab} \tensor{E}{_a^\mu}(k, \kappa) \tensor{E}{_b^\nu}(k, \kappa)
\end{align}
we immediately obtain 
\begin{align}\label{momentumspacetetrad}
	\tensor{E}{_0^0}(k, \kappa) = 1
	\qquad
	\tensor{E}{_i^\mu}(k, \kappa) = e^{\frac{k_0}{\kappa}} \delta_i^\mu.
\end{align}
Using these explicit expressions we can rewrite the action \eqref{actioncurvedms} as
\begin{align}\label{explicitdefaction}
	S_\kappa = 
	\int d\tau \,\, \dot{x}^0 k_0
	-
	e^{\frac{k_0}{\kappa}}
	\mathbf{x} \cdot \dot{\mathbf{k}}
	+ 
	N(\mathcal{C}_\kappa(k) + m^2).
\end{align}
Notice that we integrated by parts ignoring boundary terms, and furthermore in the limit $\kappa \rightarrow \infty$ the above action reduces to the action in eq. \eqref{action}. Another way of rewriting this action that is found in the literature is obtained by using the rescaling $\mathbf{k}\mapsto e^{-\frac{k_0}{\kappa}} \mathbf{k}$ so that the action becomes
\begin{align}
	S_\kappa = 
	\int d\tau \,\, \dot{x}^0 k_0
	+
	\dot{\mathbf{x}} \cdot \mathbf{k}
	+
	\mathbf{x} \mathbf{k}
	\frac{\dot{k_0}}{\kappa}
	+ 
	N(\mathcal{C}_\kappa(k) + m^2).
\end{align}
where once again we integrated by parts ignoring the boundary terms. This reformulation does not affect the physical properties of the action. 

We can now study the properties of the deformed action \eqref{explicitdefaction}. From the fact that $(p_0, \mathbf{p})$ is a Lorentz vector and $p_4$ is a Lorentz scalar, we can derive the deformed Lorentz transformations for the basis $(k_0, \mathbf{k})$. It turns out \cite{Arzano:2021scz} that their Lorentz transformations are
\begin{align}
	\delta_\lambda k_0 
	= 
	\mathbf{\lambda} \cdot \mathbf{k}
	\qquad
	\delta_\lambda \mathbf{k}_i
	=
	\mathbf{\lambda}_i
	\left(
	\frac{\kappa}{2}
	\left(
	1 - e^{-2\frac{k_0}{\kappa}}
	\right)
	+ \frac{\mathbf{k}^2}{2\kappa}
	\right)
	-
	\frac{1}{\kappa}
	\mathbf{k}_i \mathbf{\lambda} \cdot \mathbf{k}
\end{align}
and this in turn allows us to compute the Lorentz transformations of the position coordinates  that leave the deformed action invariant. After long computations, the final result is given by the following relations.
\begin{align}
	\delta_\lambda x^0 
	=
	-\mathbf{\lambda} \cdot \mathbf{x} e^{-\frac{k_0}{\kappa}}
	\qquad
	\delta_\lambda \mathbf{x}^i
	=
	-\mathbf{\lambda}^i x^0 e^{-\frac{k_0}{\kappa}}
	-\frac{1}{\kappa} 
	(\mathbf{k}^i \mathbf{\lambda} \cdot \mathbf{x} - \mathbf{\lambda}^i \mathbf{x} \cdot \mathbf{k})
\end{align}
Furthermore, the action is also invariant under the translations\footnote{Recall that momenta don't change under spacetime translations.}
\begin{align}\label{deformedtranslations}
	\delta_T x^0 = \epsilon^0
	\qquad
	\delta_T \mathbf{x}^i
	=
	\epsilon^i e^{-\frac{k_0}{\kappa}}
\end{align}
This is easily checked by noticing that, under these translations, the kinetic part of the action \eqref{explicitdefaction} acquires an extra term $\mathbf{\epsilon} \cdot \dot{\mathbf{k}}$ which is however a total derivative (and recall that for the moment we are ignoring boundary terms). This fact will be important for us when we will talk about relative locality. 

The EoM of the action \eqref{explicitdefaction}, assuming that $\mathcal{C}_\kappa(p)$ only depends on the momenta through $p_4$ and ignoring the on-shell relation $\mathcal{C}_\kappa(p) = -m^2$, are given by\footnote{The assumption that $\mathcal{C}_\kappa(p)$ depends on $p_4$ is not a restriction since a general class of models fall under this category. In any case, recall that $\kappa^2 - p_4^2 = -p_0^2 + \mathbf{p}^2$. }
\begin{align}
	\dot{k}^\mu &= 0 \\
	\dot{x}^0 
	&=
	N\frac{\partial \mathcal{C}_\kappa(p)}{\partial p_4}
	\left(
	\sinh \frac{k_0}{\kappa} - \frac{\mathbf{k}^2}{2\kappa^2} e^{\frac{k_0}{\kappa}}  
	\right) \\
	\dot{\mathbf{x}}^i 
	&=
	-N \frac{\partial \mathcal{C}_\kappa(p)}{\partial p_4}
	\frac{\mathbf{k}^i}{\kappa}
\end{align}
and these allow us to verify that the speed of light in the model described by eq. \eqref{explicitdefaction} is actually equal to one. In fact on-shell massless particles satisfy $-p_0^2 + \mathbf{p}^2 = 0$, and since $-p_0^2 +\mathbf{p}^2 + p_4^2 = \kappa^2$ and $p_4>0$, they are described by the condition $p_4 = \kappa$, which written explicitly amounts to 
\begin{align}
	\kappa \cosh \frac{k_0}{\kappa}
	-
	\frac{\mathbf{k}^2}{2\kappa} e^{\frac{k_0}{\kappa}} = \kappa.
\end{align}
Then we can easily compute that
\begin{align}\label{c=1}
	v^2 = 
	\left(
	\frac{\dot{\mathbf{x}}}{\dot{x}^0}
	\right)^2_{p_4=\kappa}
	&=
	\frac{\frac{\mathbf{k}^2}{\kappa^2}}{\left(
		\sinh \frac{k_0}{\kappa} - \frac{\mathbf{k}^2}{2\kappa^2} e^{\frac{k_0}{\kappa}}  
		\right)^2} \Bigg|_{p_4=\kappa} \nonumber \\
	&=
	\frac{2\cosh \frac{k_0}{\kappa} e^{-\frac{k_0}{\kappa}} - 2e^{-\frac{k_0}{\kappa}}}{(\sinh\frac{k_0}{\kappa} +1-\cosh\frac{k_0}{\kappa})^2}
	\Bigg|_{p_4=\kappa} \nonumber \\
	&=
	\frac{1 + e^{-2\frac{k_0}{\kappa}} - 2e^{-\frac{k_0}{\kappa}}}{\left(1 - e^{-\frac{k_0}{\kappa}}\right)^2}
	\Bigg|_{p_4=\kappa} \nonumber \\
	&= 1
\end{align}
and therefore the speed of light is energy-independent. This however does not automatically mean that photons of different energies emitted by a distant sources arrive at the same time to our detector. In fact, the phenomenon of relative locality (which we will discuss later) implies that indeed these two photons arrive at our detector at different times, but not because they have a different velocity.  

Last but not least, from the action \eqref{explicitdefaction} we can also get the symplectic form, which in turn gives us information about the Poisson brackets in our model. The starting point is the kinetic part of the action
\begin{align}
	K 
	= 
	\int d\tau \,\, \dot{x}^0 k_0
	-
	e^{\frac{k_0}{\kappa}}
	\mathbf{x} \cdot \dot{\mathbf{k}}
\end{align}
from which we can obtain the pre-symplectic form
\begin{align}
	\Theta_\kappa
	=
	\int d\tau \,\, \delta x^0 k_0
	-
	e^{\frac{k_0}{\kappa}}
	\mathbf{x} \cdot \delta \mathbf{k}.
\end{align}
Notice that in the above notation the symbol $\delta$ denotes the exterior differential in phase space. The symplectic form is then obtained by taking the exterior derivative of $\Theta_\kappa$, obtaining
\begin{align}\label{symplecticform}
	\Omega_\kappa
	=
	\delta\Theta_\kappa
	=
	\int d\tau \,\,  \delta k_0 \wedge \delta x^0
	+
	e^{\frac{k_0}{\kappa}}
	\delta \mathbf{k}_i \wedge \delta \mathbf{x}^i
	-
	\frac{\mathbf{x}^i}{\kappa}
	e^{\frac{k_0}{\kappa}}
	\delta k_0 \wedge \delta \mathbf{k}_i
	.
\end{align}
Notice that indeed we have $\delta\Omega_\kappa = 0$. The inverse of the symplectic form then gives the Poisson brackets. To understand that better consider an example inspired by \cite{Crnkovic:1986ex}. Consider a system with two degrees of freedom parametrized by the coordinates $(x^1, x^2)$ with conjugate momenta $(p^1, p^2)$. We can consider them together as a single object $Q = (p^1, p^2, x^1, x^2)$. Assume that the symplectic form is given by
\begin{align}
	\Omega = \frac{1}{2} \omega_{ab} dQ^a \wedge dQ^b = f \, dp^1 \wedge dx^1 + g \, dp^2 \wedge dx^2 + h \, dp^1 \wedge dp^2
\end{align}
where $f, g, h$ are chosen such that $d\Omega=0$. Therefore we have
\begin{align}
	\omega
	=
	\begin{pmatrix}
		0 & h & f & 0 \\
		-h & 0 & 0 & g \\
		-f & 0 & 0 & 0 \\
		0 & -g & 0 & 0
	\end{pmatrix}
\implies 
	\omega^{-1}
	=
	\begin{pmatrix}
		0 & 0 & -1/f & 0 \\
		0 & 0 & 0 & -1/g \\
		1/f & 0 & 0 & h/fg \\
		0 & 1/g & -h/fg & 0
	\end{pmatrix}.
\end{align}
We can then define the Poisson brackets between the components of $Q$ using the relation \cite{Crnkovic:1986ex}
\begin{align}
	\{A,B\} 
	= 
	(\omega^{-1})^{ab} \frac{\partial A}{\partial Q^a}
	\frac{\partial B}{\partial Q^b}.
\end{align}
One can immediately see from the above relations that 
\begin{align}
	\{x^1, p^1\} = \frac{1}{f}
	\qquad
	\{x^2, p^2\} = \frac{1}{g}
	\qquad
	\{x^1, x^2\} = \frac{h}{fg}
\end{align}
The same reasoning (with the necessary modifications) can be applied to the symplectic form in  \eqref{symplecticform}, and therefore we obtain the following Poisson brackets between the canonical variables. 
\begin{align}\label{Poissonb}
	\{x^0, p_0\} = 1
	\qquad
	\{\mathbf{x}^i, \mathbf{k}_j\} = e^{-\frac{k_0}{\kappa}}\delta^i_j
	\qquad
	\{x^0, \mathbf{x}^i\}
	=
	-\frac{1}{\kappa} \mathbf{x}^i
\end{align}
Notice that, upon quantization, the non-trivial Poisson brackets $\{x^0, \mathbf{x}^i\}$ give rise to a non-trivial commutator between spacetime coordinates, so that this model gives rise to a non-commutative spacetime. This fact can be understood as the dual counterpart to the curvature of momentum space \cite{Majid:1999tc}. This type of non-commutative spacetime is called $\kappa$-Minkowski spacetime.

\section{Group theoretical perspective} \label{GTP}

In the previous sections, we started from an adequately defined deformed action and we then derived several interesting results. In particular, we saw that the model described by the action \eqref{explicitdefaction} predicts (upon quantization) a non-commutative spacetime defined by the following Lie algebra\footnote{In the literature, one commonly uses the relation $X^0=-t$, which accounts for the difference in sign between the classical relation \eqref{Poissonb} and the algebra in eq. \eqref{kappalgebra}.}. 
\begin{align}\label{kappalgebra}
	[X^0, \mathbf{X}^i] = \frac{i}{\kappa} \mathbf{X}^i
	\qquad
	[\mathbf{X}^i, \mathbf{X}^j] = 0
\end{align}
This algebra is called the $AN(3)$ algebra, where $A$ stands for abelian (because $[\mathbf{X}^i, \mathbf{X}^j] = 0$) and $N$ for nilpotent (since $(X_i)^3 = 0$).
Notice that the objects $X^0$ and $\mathbf{X}^i$ are not the operators corresponding to spacetime positions. 
 
It is also possible to build the theory starting from the Lie algebra \eqref{kappalgebra}. It turns out that the simplest matrix representation of this algebra is (maybe unsurprisingly, given the discussion in the previous section) 5-dimensional, and it is given by the following matrices
\begin{equation}\label{rep}
	 X^0 = -\frac{i}{\kappa} \,\left(\begin{array}{ccc}
		0 & \mathbf{0} & 1 \\
		\mathbf{0} & \mathbf{0} & \mathbf{0} \\
		1 & \mathbf{0} & 0
	\end{array}\right) \quad
	\mathbf{X} = \frac{i}{\kappa} \,\left(\begin{array}{ccc}
			0 & {\mathbf{\epsilon}\,{}^T} &  0\\
			\mathbf{\epsilon} & \mathbf{0} & \mathbf{\epsilon} \\
			0 & -\mathbf{\epsilon}\,{}^T & 0
		\end{array}\right).
\end{equation}
Given a Lie algebra, we can define an element of the associated Lie group as follows\footnote{Notice that since $X^0$ and $\mathbf{X}^i$ do not comute, we have to choose an ordering for the product of the two objects $e^{i\mathbf{k}_i \mathbf{X}^i}$ and $e^{ik_0 X^0}$, and this is just a matter of convention. For example, one could have chosen to define the group element as $\tilde \Pi (k) = e^{i(\mathbf{k}_i \mathbf{X}^i - k_0 X^0)}$, and the final result would be the same as the one which we get with the convention \eqref{groupelement} but expressed in a different basis of momentum space (in this case, the basis is called `normal basis' which we will meet later).}
\begin{equation}\label{groupelement}
	\Pi(k) =e^{i\mathbf{k}_i \mathbf{X}^i} e^{ik_0 X^0}\,.
\end{equation}
and using the explicit representation \eqref{rep} one can check that
\begin{align}
	\exp(i k_0 X^0)= \left(\begin{array}{ccc}
		\cosh\frac{k_0}\kappa & \mathbf{0} & \sinh\frac{k_0}\kappa \\&&\\
		\mathbf{0} & \mathbf{1} & \mathbf{0} \\&&\\
		\sinh\frac{k_0}\kappa\; & \mathbf{0}\; & \cosh\frac{k_0}\kappa
	\end{array}\right)\, , \quad
	\exp(i \mathbf{k}_i\mathbf{X}^i) = \left(\begin{array}{ccc}
		1+\frac{\mathbf{k}^2}{2\kappa^2}\; &\; \frac{\mathbf k}\kappa \;& \; \frac{\mathbf{k}^2}{2\kappa^2}\\&&\\
		\frac{\mathbf k}\kappa & \mathbf{1} & \frac{\mathbf k}\kappa \\&&\\
		-\frac{\mathbf{k}^2}{2\kappa^2}\; & -\frac{\mathbf k}\kappa\; & 1-\frac{\mathbf{k}^2}{2\kappa^2}
	\end{array}\right)
\end{align}
and 
therefore
\begin{equation}\label{groupelement2}
	\Pi(k)   =\left(\begin{array}{ccc}
		\frac{ \bar p_4}\kappa \;&\;  \frac{\mathbf k}\kappa \; &\;
		\frac{  p_0}\kappa\\&&\\
		\frac{\mathbf p}\kappa  & \mathbf{1} & \frac{\mathbf p}\kappa  \\&&\\
		\frac{\bar p_0}\kappa\; & -\frac{\mathbf k}\kappa\; &
		\frac{  p_4}\kappa
	\end{array}\right)\, .
\end{equation}
where $p_0, \mathbf{p}, p_4$ are defined in eq. \eqref{5dcoord} and 
\begin{align}
	\bar{p}_0(k_0, \mathbf{k}) &= \kappa \sinh \frac{k_0}{\kappa}
	-
	\frac{\mathbf{k}^2}{2\kappa} e^{\frac{k_0}{\kappa}} \nonumber  \\
	\bar{p}_4(k_0, \mathbf{k})
	&=
	\kappa \cosh \frac{k_0}{\kappa}
	+
	\frac{\mathbf{k}^2}{2\kappa} e^{\frac{k_0}{\kappa}} \nonumber 
\end{align}
Notice that $\Pi(k)$ is completely determined once $p_0, \mathbf{p}, p_4$ are known\footnote{One can easily express also $\bar{p}_0, \bar{p}_4, \mathbf{k}$ as a function of $p_0, \mathbf{p}, p_4$. For example, one can check that $\bar{p}_0 = p_0 - \frac{\mathbf{p}^2}{p_0+p_4}$.}. Now that we have a generic group element, we can also get a description of the group manifold. In order to get it, we act with all possible group elements on a fixed vector $\mathcal{O}$ in the 5-dimensional vector space on which $\Pi(k)$ acts\footnote{More technically, one can say that the group acts transitively on $\mathcal{O}$, i.e. the orbit of any point in the group is the whole group.}. We choose $\mathcal{O} = (0,0,0,0,\kappa)^T$, which represents the momentum space origin and can physically be interpreted as a point of zero energy and zero momentum (notice that this point is Lorentz invariant). As a result, for any fixed $\Pi(k)$, the object $\Pi(k) \mathcal{O}$ is a point in the 5-dimensional momentum space with coordinates $(p_0, \mathbf{p}, p_4)$ which is in one-to-one correspondence with the group element $\Pi(k)$. However, we already saw that the coordinates $(p_0, \mathbf{p}, p_4)$ describe a 4-dimensional (submanifold of the) de Sitter space, which is therefore our group manifold.

\section{Relative locality and interactions}

\begin{figure}
    \centering
    \begin{tikzpicture}
		
		\draw[color=black, thick] (0,0) -- (1.72,-1);
		\draw[color=black, thick] (0,0) -- (-1.72,-1);
		\draw[color=black, thick] (0,0) -- (0,2);
		
		\draw[color=black, thick, -<] (0,0) -- (0.86,-0.5);
		\draw[color=black, thick, -<] (0,0) -- (-0.86,-0.5);
		\draw[color=black, thick, ->] (0,0) -- (0,1);
		
		\filldraw[color=black] (0,0) circle (0.8mm);
		
		\draw[color=blue, ->] (-2,0) node[label={[xshift=-0.3cm, yshift=-0.2cm]$A$}]{} -- (-1.5,0);
		\draw[color=blue, ->] (-2,0) -- (-2,0.5);
		\draw[color=blue, ->] (-2,0) -- (-2.3,-0.3);

	\end{tikzpicture}
    \caption{Local interaction with respect to the observer $A$ \\ local with the interaction}
    \label{FIG1}
\end{figure}
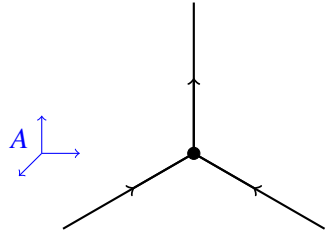

We now discuss relative locality, an unexpected feature of theories with nontrivial momentum space geometry \cite{Amelino-Camelia:2011lvm}, \cite{Amelino-Camelia:2011hjg}. Let us use as a starting point a local interaction of point-like objects, like the one depicted in the Figure \ref{FIG1}.

This type of graph of an interaction has actually a wider scope than just a representation of point-like particles, since the same structure can be found for Feynman diagrams in QFT. Furthermore, the only relativistic invariant potential in SR is the Dirac delta potential, which once again describes a contact interaction. Because of this universality, an interaction like this is usually referred to as an event, and indeed in canonical GR one can uniquely define a spacetime point by assigning an event to it. In fact, an event is an absolute concept in GR abd QFT since, although different observers can have different measurement of the scattering properties (such as the momenta of the particles involved), they all agree on the fact that the scattering happened in the first place, and that it happened locally at the same spacetime point\footnote{Of course, general covariance in GR only describes our freedom to give this spacetime point any name we like, i.e. any numerical value in terms of our favourite coordinates, but the construction of a spacetime point in terms of event is not influenced by this.}. Incidentally, this is also one of the problematic points of a potential theory of quantum gravity, since such contact interaction between gravitons leads to renormalization issues \cite{Goroff:1985th}.

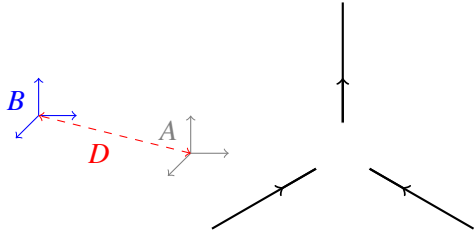
\begin{figure}
    \centering
	\begin{tikzpicture}
		
		\draw[color=black, thick] (0,0) -- (1.72,-1);
		\draw[color=black, thick] (0,0) -- (-1.72,-1);
		\draw[color=black, thick] (0,0) -- (0,2);
		
		\draw[color=black, thick, -<] (0,0) -- (0.86,-0.5);
		\draw[color=black, thick, -<] (0,0) -- (-0.86,-0.5);
		\draw[color=black, thick, ->] (0,0) -- (0,1);
		
		\filldraw[color=white] (0,0) circle (4mm);
		
		\draw[color=gray, ->] (-2,0) node[label={[xshift=-0.3cm, yshift=-0.1cm]$A$}]{} -- (-1.5,0);
		\draw[color=gray, ->] (-2,0) -- (-2,0.5);
		\draw[color=gray, ->] (-2,0) -- (-2.3,-0.3);
		
		\draw[color=blue, ->] (-4,0.5) node[label={[xshift=-0.3cm, yshift=-0.2cm]$B$}]{} -- (-3.5,0.5);
		\draw[color=blue, ->] (-4,0.5) -- (-4,1);
		\draw[color=blue, ->] (-4,0.5) -- (-4.3,0.2);
		
		\draw[color=red, dashed, <->] (-2,0) -- (-4,0.5) node[label={[xshift=0.8cm, yshift=-0.9cm]$D$}]{};
		
	\end{tikzpicture}
	\caption{\footnotesize{Same interaction as before, but seen from \\ $\qquad$ an observer $B$ distant from $A$}}
    \label{FIG2}
\end{figure}

The universality of contact interactions make them the ideal starting point for our considerations on relative locality.   As a matter of fact, we already encountered one of the key aspects of relative locality when we wrote down the transformations our coordinates under translations in eq. \eqref{deformedtranslations} which leave the deformed action invariant (which we reproduce here for simplicity).
\begin{align}
	\delta_T x^0 = \epsilon^0
	\qquad
	\delta_T \mathbf{x}^i
	=
	\epsilon^i e^{-\frac{k_0}{\kappa}}
\end{align}
Contrary to SR, translations act in different way on particles of different energies, and therefore an observer $B$ translated by a distance $D$ with respect to an observer $A$ near the interaction will see a different interaction, represented in the Figure \ref{FIG2}.

Therefore, in this theory locality is not observer independent. An event that is local for one observer, Figure \ref{FIG1}, is not local to another, Figure \ref{FIG2}. There is now the issue of defining spacetime, because the canonical description of a spacetime point as event does not hold anymore in our context. Indeed in Special Relativity a spacetime point is defined  as an event that happens at this point, ie., by some physical local process taking place there. The only elementary process one can think of is the interaction as depicted in Figure \ref{FIG1}. In Special Relativity such interaction defines an event and the spacetime point because all the observers agree that the interaction is local so the definition of the spacetime point as the point where the interaction takes place is observer-independent. In the case of relative locality this is not the case, and the points are sharply defined by interactions only to the observers who are local to the interaction point.

One of the important properties of the interaction point is that some quantities, for example momenta, are conserved in the course of the interaction. It is therefore reasonable to start our investigation by listing all the properties that an interaction should have. 

\begin{figure}
    \centering
	\begin{tikzpicture}
		
		\draw[color=black, thick] (0,0) -- (1.72,-1);
		\draw[color=black, thick] (0,0) -- (-1.72,-1);
		\draw[color=black, thick] (0,0) -- (0,2);
		
		\draw[color=black, thick, -<] (0,0) -- (0.86,-0.5) node[label={[xshift=0cm, yshift=0cm]$p$}]{};
		\draw[color=black, thick, -<] (0,0) -- (-0.86,-0.5) node[label={[xshift=0cm, yshift=0cm]$q$}]{};
		\draw[color=black, thick, ->] (0,0) -- (0,1) node[label={[xshift=0.3cm, yshift=0cm]$r$}]{};
		
		\filldraw[color=black, fill=white, thick] (0,0) circle (4mm);

	\end{tikzpicture}
	\caption{3-valent vertex}
	\label{FIG3}
\end{figure}
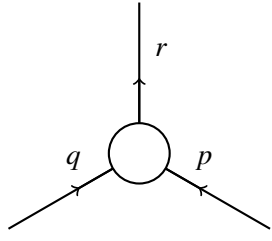

First of all, in a trivial (2-valent) vertex with one particle coming in, nothing happening in between, and one particle going out, if the initial momentum is $p$ then also the final momentum should be $p$ (by conservation of energy and momentum). Mathematically we can write that 
\begin{align}
	0\oplus p = p \oplus 0 = p
\end{align}
where $\oplus$ is some abstract composition of momenta. Of course, since we have this new composition law, we also expect to be able to have particles with momenta $S(p)$ which is opposite to $p$, i.e. such that
\begin{align}
	p \oplus S(p) = S(p) \oplus p = 0.
\end{align}
Notice that at this point we are still not assuming associativity, so we have the structure of a quasigroup. Consider now  a 3-valent vertex like the one in the Figure \ref{FIG3}.
In this case we have two particles coming in and one particle coming out, and we expect that the following relations hold.
\begin{align}\label{defmomcons}
	p\oplus q = r
	\qquad
	S(r) \oplus (p\oplus q) =  (p\oplus q) \oplus S(r) = 0
\end{align}
Now impose the condition that the above relations are Poincar\'e covariant in an appropriate, deformed sense. Considering boosts (because spacetime rotations are assumed to act trivially) and using the more abstract notation used in eq. \eqref{abstractvariations} we impose the covariance of the relations \eqref{defmomcons}
\begin{align}
	N_i \triangleright (p\oplus q)
	\equiv
	\sum_i
	(\{\mathcal{P}_i^1 \triangleright p \}
	+
	\{\mathcal{P}_i^2 \triangleright q \})
	=
	N_i \triangleright r
\end{align}
where we made use of the so-called Sweedler notation, and where $\mathcal{P}_i^1, \mathcal{P}_i^2$ are appropriate generators of the Poincar\'e algebra. This consistency condition ensures that the Poincar\'e generators form a Hopf algebra (which is a generalization of Lie algebras), but we will not go into more details in this direction. In the undeformed case, after a boost one would have $p_\mu \mapsto N_i\triangleright p_\mu = p_\mu + \delta_i p_\mu$ and $q\mu \mapsto N_i\triangleright q_\mu = q_\mu + \delta_i q_\mu$, and therefore
\begin{align}
	p_\mu + q_\mu 
	\mapsto
	N_i\triangleright p_\mu 
	+
	N_i\triangleright q_\mu
	=
	p_\mu + q_\mu 
	+
	\delta_i p_\mu + \delta_i q_\mu 
	= 
	r_\mu+\delta_i r_\mu
	=
	N_i \triangleright r_\mu
\end{align}
which therefore implies 
\begin{align}
	\mathcal{P}_1^1 = \mathcal{P}_1^2 = 1
	\qquad
	\mathcal{P}_2^1 = \mathcal{P}_2^2 = \delta_i.
\end{align}
Notice that the action of $N_i$ is linear because in the undeformed case we don't have a scale, which would make it possible for nonlinearities to appear.

Now we have all the properties that we would like to have for the deformed composition of momenta, and we only need to specify explicitly what this composition rule actually is. In order to do it, we come back to the group theoretical perspective. Given two elements $\Pi(k), \Pi(l)$ of the group, we know that also their product $\Pi(k)\Pi(l)$ will be an element of the group, and we can use this group multiplication to define the addition of momenta. 
\begin{align}
	e^{i(\mathbf{k}\oplus \mathbf{l})_i \mathbf{X}^i} e^{i(k_0\oplus l_0) X^0}
	=
	\Pi(k \oplus l)
	:=
	\Pi(k)\Pi(l)
	=
	e^{i\mathbf{k}_i \mathbf{X}^i} e^{ik_0 X^0}
	e^{i\mathbf{l}_i \mathbf{X}^i} e^{il_0 X^0}
\end{align}
The above product can be easily preformed using the Baker–Campbell–Hausdorff formula or the explicit expression in eq. \eqref{groupelement2}, obtaining
\begin{align}
	e^{i(\mathbf{k}\oplus \mathbf{l})_i \mathbf{X}^i} e^{i(k_0\oplus l_0) X^0}
	=
	e^{i\left(\mathbf{k}_i + e^{-k_0/\kappa} \mathbf{l}_i\right) \mathbf{X}^i} e^{i(k_0 + l_0) X^0}
\end{align}
which means that 
\begin{align}
	(\mathbf{k}\oplus \mathbf{l})_i
	&=
	\mathbf{k}_i + e^{-k_0/\kappa} \mathbf{l}_i \label{comprule1}\\
	(k_0 \oplus l_0)
	&=
	k_0 + l_0 \label{comprule2}
\end{align}
Notice that the energies sum like in the undeformed case, but the composition law for spatial momenta is not linear due to the factor $e^{-\frac{k_0}{\kappa}}$. Furthermore, this composition law is non-abelian and associative (due to the group properties). 

More in general, one can view the deformed composition law for momenta more geometrically by relating it to the properties of momentum space. In fact, one can write 
\begin{align}\label{connectiondef}
	(p\oplus q)_\mu
	=
	p_\mu + q_\mu
	-
	\tensor{\Gamma}{_\mu^{\alpha\beta}} p_\alpha q_\beta
	+
	\dots
	\qquad
	\tensor{\Gamma}{_\mu^{\alpha\beta}}
	=
	-
	\frac{\partial^2 (p\oplus q)_\mu}{\partial p_\alpha \partial q_\beta} \Bigg|_{p=q=0}.
\end{align}
Recall that $p,q$ from a geometrical point of view describe the coordinates of some points in the 4-dimensional de Sitter space, and $(p\oplus q)$ is a different point in momentum space, and to go from one point to another in a curved momentum space one really needs to define a parallel transport, and therefore a connection. Notice however that the connection in  \eqref{connectiondef} is associated with the $\oplus$ operation which involves two points, and it is therefore not immediately related to the metric of momentum space, which only involves one (see for example  \eqref{momentumspacemetric}). Indeed, having defined a connection in  \eqref{connectiondef}, and since we already know the metric in momentum space, it is now possible to see that the following hold:
\begin{itemize}
	\item The connection $\Gamma$ is torsion-free iff the composition rule $\oplus$ is symmetric;
	\item The connection $\Gamma$ has curvature (i.e. the curvature associated with the connection $\Gamma$ is non-zero) iff the composition rule $\oplus$ is not associative;
	\item The connection $\Gamma$ is not in general metric since $\nabla_\Gamma G^{\mu\nu} \neq 0$.
\end{itemize}
Referring back to the composition rules in  \eqref{comprule1}, \eqref{comprule2}, our connection would therefore be with torsion, zero curvature, and non-metric. 

We now want to describe the interaction at the level of the action, i.e. we want to generalize the single particle action in \eqref{actioncurvedms} to the case of many interacting particles (for simplicity, here we consider only scalar particles without spin). To start with we notice that an interacting particle's worldline is semi-infinite, and in particular for incoming particles $\tau \in (-\infty, 0)$ and for outgoing particles $\tau \in (0, +\infty)$ (here we are choosing a $\tau$ such that the interaction happens at $\tau = 0$). Therefore the single incoming particle action is first rewritten as
\begin{align}\label{actioncurvedmsincoming}
	S_\kappa = 
	-\int_{-\infty}^0 d\tau \,\, x^a \tensor{E}{_a^\mu}(k, \kappa) \dot{k}_\mu - N(\mathcal{C}_\kappa(k) + m^2).
\end{align}
and similarly for outgoing particles, so that the total free Lagrangian of incoming and outgoing particles is now given by
\begin{align}\label{freeaction}
	S_\kappa^{\text{free}} = 
	-\sum_j
	\int_{\text{in}/\text{out}} d\tau \,\, x^a_j \tensor{E}{_a^\mu}(k^j, \kappa) \dot{k}_\mu^j - N^j(\mathcal{C}_\kappa^j(k^j) + m^2_j).
\end{align}
For scalar particles, the only quantity which is conserved in an interaction is momentum, so we can define the interaction term in the action as
\begin{align}\label{intaction}
	S_\kappa^{\text{int}}
	:=
	z^\mu 
	(k^1 \oplus k^2 \oplus \dots )_\mu
\end{align}
where $z^\mu$ is just a Lagrange multiplier enforcing momentum conservation at the vertex. Notice that in this case the importance of having the time derivative on $k$ and not on $x$ in the free action \eqref{freeaction} is highlighted even more. In fact, while  previously we didn't have a boundary so that we could have in principle integrated by parts, now we do have a boundary, and therefore the choice of whether to use $\dot{x}^\mu k_\mu$ or $x^\mu \dot{k}_\mu$ in the deformed action is crucial, since the two are not equivalent. 

The equations of motion following from the action $S_\kappa^{\text{free}} + S_\kappa^{\text{int}}$ will now have both a bulk contribution and a surface one. The bulk equations of motion  are given by
\begin{align}
	&\dot{k}_\mu^j = 0 \nonumber \\
	&\dot{x}^\mu_j = N^j  \tensor{E}{_a^\mu}(k^j, \kappa) \frac{\partial \mathcal{C}_\kappa^j(k^j)}{\partial k_\mu^j} \\
	&\mathcal{C}_\kappa^j(k^j) + m^2 = 0
\end{align}
and the equations of motion at the interaction point has the form 
\begin{align}
	x^a_j(0)
	=
	z^\nu
	\tensor{E}{^a_\mu}
	\frac{\partial}{\partial k^j_\mu}
	(k^1 \oplus k^2 \oplus \dots )_\nu.
\end{align}
The boundary equations of motion  highlight again the effect of relative locality already depicted in Figure \ref{FIG2}.  Let us look at translations for simplicity (there is a similar effect  for Lorentz transformations). We have
\begin{align}\label{rellocvertex}
	\delta x^a_j(0)
	=
	\delta z^\nu
	\tensor{E}{^a_\mu}
	\frac{\partial}{\partial k^j_\mu}
	(k^1 \oplus k^2 \oplus \dots )_\nu.
\end{align}
Notice that since $\delta x^a_j(0)\neq \delta x^a_i(0)$ for $i\neq j$, it follows from \eqref{rellocvertex} that translations are momentum dependent, which means that particles of different momenta will be translated differently, which is indeed the behaviour shown in Figure \ref{FIG2}. The only way to avoid translations being momentum dependent would be to have a flat momentum space, which translates to a trivial momentum space tetrad and a linear composition of momenta, in which case  \eqref{rellocvertex} reduces to $\delta x^a_j(0) = \delta z^a$, so that the endpoints of all the worldlines translate by the same amount. Of course, in this case the interaction event behaves as in canonical SR, with the standard locality. 
One could also think to extend this reasoning to many vertices and to loops, but the situation in this case is still not completely clear (especially when loops are involved). 

Finally, as a final demonstration of phenomena related to relative locality, we show how two photons with different energies emitted simultaneously (with respect to an observer nearby the source) in a distant objects can be detected on Earth as having a time delay, despite the fact that both travel at the same speed $c=1$ (as shown in  \eqref{c=1}). The reason why this is possible is of course related to the fact that something that can be local for a distant observer, because of relative locality it is not local anymore from the point of view of our detectors on Earth. Let us consider the situation (depicted below) of one high-energy photon with momentum $(k_0^2, 0,0,\mathbf{k}_z^2)$ and a low-energy one with momentum $(k_0^1, 0,0,\mathbf{k}_z^1)$ emitted at the same time locally to an observer $A$ distant $d$ from Earth.
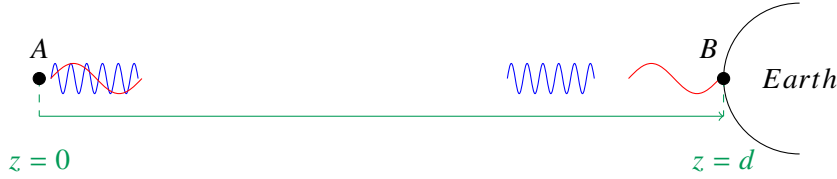
\begin{center}
	\begin{tikzpicture}
		
		\draw[color=black] (10,-1) node[label={[xshift=0cm, yshift=0.6cm]$Earth$}]{} arc (270:90:1cm);
		
		\draw[x=0.052cm,y=0.2cm, blue] 
		(3,0) sin (4,1) cos (5,0) sin (6,-1) cos (7,0)
		sin (8,1) cos (9,0) sin (10,-1) cos (11,0) sin (12,1) cos (13,0) sin (14,-1) cos (15,0) sin (16,1) cos (17,0) sin (18,-1) cos (19,0) sin (20,1) cos (21,0) sin (22,-1) cos (23,0) sin (24,1) cos (25,0);
		\draw[x=0.3cm,y=0.2cm, red] 
		(0.5,0) sin (1.5,1) cos (2.5,0) sin (3.5,-1) cos (4.5,0);
		
		\begin{scope}[shift={(6,0)}]
			\draw[x=0.052cm,y=0.2cm, blue] 
			(3,0) sin (4,1) cos (5,0) sin (6,-1) cos (7,0)
			sin (8,1) cos (9,0) sin (10,-1) cos (11,0) sin (12,1) cos (13,0) sin (14,-1) cos (15,0) sin (16,1) cos (17,0) sin (18,-1) cos (19,0) sin (20,1) cos (21,0) sin (22,-1) cos (23,0) sin (24,1) cos (25,0);
		\end{scope}
		\begin{scope}[shift={(7.6,0)}]
		\draw[x=0.3cm,y=0.2cm, red] 
		(0.5,0) sin (1.5,1) cos (2.5,0) sin (3.5,-1) cos (4.5,0);	
		\end{scope}
	
		\draw[color=ForestGreen, ->] (0,-0.5) -- (9,-0.5) node[label={[xshift=-9cm, yshift=-1cm]$z=0$}]{} node[label={[xshift=0cm, yshift=-1cm]$z=d$}]{};
		\draw[color=ForestGreen, dashed] (0,0) -- (0,-0.5);
		\draw[color=ForestGreen, dashed] (9,0) -- (9,-0.5);
		
		\filldraw[color=black] (0,0) node[label={[xshift=0cm, yshift=0cm]$A$}]{} circle (0.8mm);
		
		\filldraw[color=black] (9,0) node[label={[xshift=-0.2cm, yshift=0cm]$B$}]{} circle (0.8mm);
		
	\end{tikzpicture}
\end{center}
Since $c^2 = \left(\frac{\dot{\mathbf{X}}^z}{\dot{X}^0}\right)^2 = 1$, the equations of motion for the low-energy photon will be given by
\begin{align}
	\dot{X}_1^0 = A(k^1) := A_1
	\qquad
	\dot{\mathbf{X}}_1^z = A(k^1) := A_1
\end{align}
which can immediately be integrated to give
\begin{align}\label{solAlow}
	X^0_1 = A_1 \tau
	\qquad
	\mathbf{X}^z_1 = A_1 \tau
\end{align}
where there is no additional constant $C$ because we are at the moment in the reference frame $A$. In the same way, for the energetic photon we get
\begin{align}\label{solAhigh}
	X^0_2 = A_2 \tau
	\qquad
	\mathbf{X}^z_2 = A_2 \tau
\end{align}
Now we have to translate the solutions \eqref{solAlow}, \eqref{solAhigh} to the observer $B$, so that we can predict what they will measure. To do so, assume that $k_0^1$ is small enough that $e^{-\frac{k_0}{\kappa}} \approx 1$, so that for the low-energy photon the translation acts in the standard way so that for $B$ we have.
\begin{align}\label{solBlow}
	X^0_1 = A_1 \tau
	\qquad
	\mathbf{X}^z_1 = A_1 \tau - d
\end{align}
If instead $k_0^2$ is high enough, then we cannot neglect the factor $e^{-\frac{k_0^2}{\kappa}}$ coming from the shift of the second photon. Therefore from $B$ point of view we would have
\begin{align}\label{solBhigh}
	X^0_2 = A_2 \tau
	\qquad
	\mathbf{X}^z_2 = A_2 \tau - e^{-\frac{k_0^2}{\kappa}} d.
\end{align}
Now notice that the low-energy photon reaches us when $\tau = \frac{d}{A_1}$, and at this instant the high energy photon is still distant
\begin{align}
	\mathbf{X}^z_2(\tau = d/A_1)
	=
	d
	\left(
	\frac{A_2}{A_1}
	-
	e^{-\frac{k_0^2}{\kappa}}
	\right)
\end{align}
from $B$. In general this distance is different from zero, which provides the anticipated difference in arrival times. 

Notice one subtlety of this example. In the whole discussion about relative locality, interactions play a crucial role. Indeed, relative locality can be understood as the fact that an interaction which is local for an observer is not local for some other one. However, in the above example we have non-interacting photons, which are created in different processes and measured with different apparatuses, and yet we still have effects due to relative locality. The key point is that, although the photons are not interacting between themselves, they were (individually) created by some interaction between other particles\footnote{For example, the low energy photon could have been created due to a low energy scattering between charged particles, and the high energy photon by a positron-electron annihilation}. These interactions are both local for $A$ (since both photons are created locally to $A$), but are not local for $B$. However, the amount by which these interactions are non-local for $B$ depends on the energy of the photons. In particular, the event which generated the low-energy photon will seem approximately local also for $B$, while the one which generated the energetic photon will not be local for $B$. The situation can be schematically represented by the picture below.
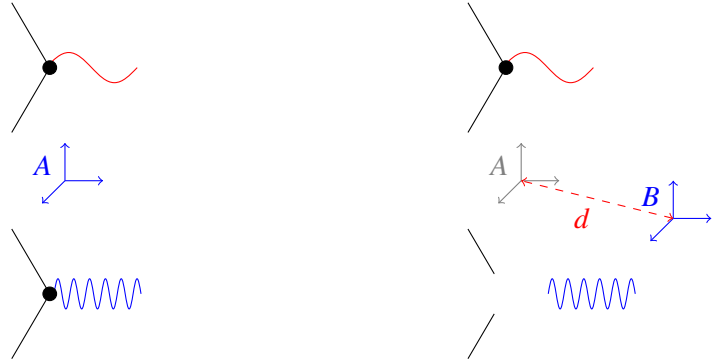
\begin{center}
	\begin{tikzpicture}

		\begin{scope}[shift={(0.1,0)}]
			\draw[x=0.052cm,y=0.2cm, blue] 
			(3,0) sin (4,1) cos (5,0) sin (6,-1) cos (7,0)
			sin (8,1) cos (9,0) sin (10,-1) cos (11,0) sin (12,1) cos (13,0) sin (14,-1) cos (15,0) sin (16,1) cos (17,0) sin (18,-1) cos (19,0) sin (20,1) cos (21,0) sin (22,-1) cos (23,0) sin (24,1) cos (25,0);
		\end{scope}
		\begin{scope}[shift={(0,3)}]
			\draw[x=0.3cm,y=0.2cm, red] 
			(0.5,0) sin (1.5,1) cos (2.5,0) sin (3.5,-1) cos (4.5,0);
		\end{scope}
		\begin{scope}[shift={(0.2,0)}]
			\filldraw[color=black] (0,0) circle (0.9mm);
			
			\draw[color=black] (-0.5,-0.86) -- (0,0) -- (-0.5, 0.86);
		\end{scope}
		\begin{scope}[shift={(0.2,3)}]
			\filldraw[color=black] (0,0) circle (0.9mm);
		
			\draw[color=black] (-0.5,-0.86) -- (0,0) -- (-0.5, 0.86);
		\end{scope}
	\begin{scope}[shift={(4.4,1)}]
		\draw[color=blue, ->] (-4,0.5) node[label={[xshift=-0.3cm, yshift=-0.2cm]$A$}]{} -- (-3.5,0.5);
		\draw[color=blue, ->] (-4,0.5) -- (-4,1);
		\draw[color=blue, ->] (-4,0.5) -- (-4.3,0.2);
	\end{scope}

		\begin{scope}[shift={(6.6,0)}]
			\draw[x=0.052cm,y=0.2cm, blue] 
			(3,0) sin (4,1) cos (5,0) sin (6,-1) cos (7,0)
			sin (8,1) cos (9,0) sin (10,-1) cos (11,0) sin (12,1) cos (13,0) sin (14,-1) cos (15,0) sin (16,1) cos (17,0) sin (18,-1) cos (19,0) sin (20,1) cos (21,0) sin (22,-1) cos (23,0) sin (24,1) cos (25,0);
		\end{scope}
		\begin{scope}[shift={(6,3)}]
			\draw[x=0.3cm,y=0.2cm, red] 
			(0.5,0) sin (1.5,1) cos (2.5,0) sin (3.5,-1) cos (4.5,0);
		\end{scope}
		\begin{scope}[shift={(6.2,0)}]
			\draw[color=black] (-0.5,-0.86) -- (0,0) -- (-0.5, 0.86);
			
			\filldraw[color=white] (0,0) circle (3mm);
		\end{scope}
		\begin{scope}[shift={(6.2,3)}]
			\filldraw[color=black] (0,0) circle (0.9mm);
			
			\draw[color=black] (-0.5,-0.86) -- (0,0) -- (-0.5, 0.86);
		\end{scope}
		\begin{scope}[shift={(10.4,1)}]
			\draw[color=blue, ->] (-2,0) node[label={[xshift=-0.3cm, yshift=-0.1cm]$B$}]{} -- (-1.5,0);
			\draw[color=blue, ->] (-2,0) -- (-2,0.5);
			\draw[color=blue, ->] (-2,0) -- (-2.3,-0.3);
			
			\draw[color=gray, ->] (-4,0.5) node[label={[xshift=-0.3cm, yshift=-0.2cm]$A$}]{} -- (-3.5,0.5);
			\draw[color=gray, ->] (-4,0.5) -- (-4,1);
			\draw[color=gray, ->] (-4,0.5) -- (-4.3,0.2);
			
			\draw[color=red, dashed, <->] (-2,0) -- (-4,0.5) node[label={[xshift=0.8cm, yshift=-0.9cm]$d$}]{};
		\end{scope}
		
	\end{tikzpicture}
\end{center}

\section{The soccer ball problem}

The soccer ball problem for theories with deformed dispersion relation and/or modified composition laws has been formulated almost immediately after deformed theories were first formulated\footnote{According to J.\ Lukierski it was already in early 1990th when during one of his seminars I. Bia\l{}ynicki-Birula objected that if the deformed dispersion relation was universally valid for macroscopic bodies it would contradict every day observations. Unfortunately, part of the community still keeps thinking that the soccer ball problem is an unsolved paradox.}. The problem can be roughly summarized as follows. In the deformed context we deal with deformed dispersion relations like 
\begin{align}\label{defdisp}
	\mathcal{C}_\kappa(k)
	=
	-k_0^2 + \mathbf{k}^2 
	+
	\frac{1}{\kappa}
	\Delta^{\mu\nu\rho}
	k_\mu k_\nu k_\rho
	+
	\dots
	=
	-m^2
\end{align}
and composition rules like the one in eq. \eqref{connectiondef}. All of them depend on the scale $\kappa$ which has been central in our discussion. Since the deformation in DSR models is assumed to be of quantum gravitational origin, one usually expects $\kappa$ to be of the order of the Planck mass, which is around $10^{19} \, GeV$. Furthermore, when one deals about composition of momenta and scattering of particles, one usually has fundamental particles in mind (like electron, photons and so on) for which one usually has $E/\kappa <<1$, where $E$ is the particle energy. One then can argue that since $\kappa$ is universal, the same composition rules and deformed dispersion relation should hold no matter how many particles are involved. In particular, we could also consider a  soccer ball with all its particles, and for this object,  the effects of the deformed composition rules are at the first sight very big. Indeed, although every single particle has the energy by many orders of magnitudes smaller than the Planck energy (i.e. $E/\kappa <<1$), there are so many particles that the energy of the system is much higher than $\kappa$, so that the deformations used in DSR should have a large macroscopic effect. The problem now arises because such large deformations are obviously not observed in macroscopic objects. 
In other words, how can composition rules and dispersion relations be universally deformed, and yet a soccer ball, for which $p/\kappa \approx 10^8$ (and therefore the contributions due to deformations should be enormous) behaves classically?

To describe the solution to this problem it is convenient to use the  so-called normal basis \cite{Amelino-Camelia:2011dwc} (since the choice of basis, being a choice of coordinates, cannot have any physical effect, it is just a technical simplification). In fact, mathematically, the soccer ball problem relies on the fact that, once we generalize  \eqref{connectiondef} to the sum of $N$ momenta, the number of quadratic contributions grows like $N^2$. A solution of the soccer problem would therefore amount to some way of dealing with this growth. Notice that the qualitative argument on the growth of terms like $N^2$ relies on  \eqref{connectiondef}, which however requires an explicit choice of basis to be made. To deal with the soccer ball problem, we pick the so called normal basis. We already encountered normal coordinates $\tilde{p}_\mu$ when talking about the convention used to write eq. \eqref{groupelement} and they are defined by
\begin{align}
	e^{i\mathbf{k}_i \mathbf{X}^i} e^{ik_0 X^0}
	=
	e^{i \tilde{p}_\mu X^\mu}.
\end{align}
Because of their definition, since $[X^\mu, X^\mu]=0$, they satisfy the relation
\begin{align}
	(\tilde{p}\oplus \tilde{p})_\mu = 2\tilde{p}_\mu
\end{align}
 To give more explicit expressions, proceeding as in section \ref{GTP}, one can show that \cite{Kowalski-Glikman:2017ifs}
\begin{align}
	\tilde{p}_0
	=
	k_0
	\qquad
	\tilde{\mathbf{p}}_i
	=
	\mathbf{k}_i 
	\frac{\frac{k_0}{\kappa}}{1- e^{-\frac{k_0}{\kappa}}}
\end{align}
with the composition rules
\begin{align}
	(\tilde{p}\oplus \tilde{q})_0 = \tilde{p}_0 + \tilde{q}_0
	\qquad
	(\tilde{\mathbf{p}}\oplus \tilde{\mathbf{q}})_i
	=
	\left(
	\tilde{\mathbf{p}}_i \frac{1}{f(\tilde{p}_0)}
	+
	\tilde{\mathbf{q}}_i \frac{e^{-\frac{\tilde{p_0}}{\kappa}}}{f(\tilde{q}_0)}
	\right)
	f(\tilde{p}_0 + \tilde{q}_0)
\end{align}
where
\begin{align}
	f(\tilde{p}_0)
	=
	\frac{\tilde{p_0}}{\kappa}
	\frac{1}{1-e^{-\frac{\tilde{p}_0}{\kappa}}}
\end{align}
Using these one can indeed verify that $(\tilde{p}\oplus \tilde{p})_0 = 2\tilde{p}_0$ and 
\begin{align}
	(\tilde{\mathbf{p}}\oplus \tilde{\mathbf{p}})_i
	=
	\frac{f(2\tilde{p}_0)}{f(\tilde{p}_0)}
	\left(
	1 + e^{-\frac{\tilde{p}_0}{\kappa}}
	\right)
	\tilde{\mathbf{p}}_i
	=
	2 \tilde{\mathbf{p}}_i.
\end{align}
In the normal basis, if we consider the ball to be formed by $N$ particles\footnote{This of course now bear the question of what are the fundamental constituents of the ball, because depending on what they are the number $N$ changes. However, even restricting our attention to atoms, then $N$ would be big enough to make our reasoning valid. Of course, if one chooses then to consider electrons, protons and so on, then one would get a bigger $N$.} each with momentum $\tilde{p}$, then the total momentum of the ball would just be $(\tilde{p}\oplus(\tilde{p}\oplus \dots)) = N \tilde{p} := \tilde{P}$, without any contribution coming from deformations. At the same time, the dispersion relation for the whole ball can be easily obtained from eq. \eqref{defdisp} by summing the individual dispersion relations of the individual particles, obtaining
\begin{align}
	-P_0^2 + \mathbf{P}^2 
	+
	\frac{1}{N\kappa}
	\Delta^{\mu\nu\rho}
	P_\mu P_\nu P_\rho
	+
	\dots
	=
	-N^2m^2
	=
	-M^2
\end{align}
where $M=Nm$ is the total mass of the ball. We see that the deviation from the undeformed dispersion relation is of the order $\frac{P}{N\kappa}$, i.e. it is negligibly small.
Notice that we chose the normal basis because the solution to the soccer ball problem is particularly transparent, but if one prefers one can chose any other coordinates and work out explicitly the whole argument with them. A similar procedure shows that for scattering of macroscopic bodies one can get 
\begin{align}
	P^{in}_1 + P^{in}_2 = P^{out}_1 + P^{out}_2 + O\left(\frac{1}{N\kappa}\right)
\end{align}
This shows that for macroscopic bodies the effective deformation parameter is not $\kappa$ but $N\kappa$, which is a number at least by the factor of order of $10^{23}$ times larger. The reader is encouraged to investigate how the argument would change if one replaces the normal coordinates in momentum space with different ones.

There is however yet another potential problem \cite{Hossenfelder:2012vk} that needs to be solved when one allows for fluctuations around an average value, i.e. $k = \bar{k} + \delta k$ with the average fluctuation equal zero $\langle\delta k \rangle = 0$. The issue is that, even though $\langle\delta k \rangle = 0$, the amplitude of the fluctuation must be small too. If this amplitude were to be relevant, the soccer ball would be macroscopically fluctuating around its classical trajectory, which is once again a behaviour which is not observed in reality. Not all deformed theories pass this test since some indeed predict macroscopic fluctuations of large bodies, and indeed this condition restrict the possible geometries of momentum space \cite{Amelino-Camelia:2013zja}.

\section{Relations between deformations and quantum gravity}

In this concluding section we will briefly discuss  the relation between the deformation of relativistic symmetries discussed up to now, and the problem of quantum gravity. The starting point is given by general relativity in $2+1$ dimensions.

In 3-dimensional gravity there are no local degrees of freedom (i.e. no local gravitational degrees of freedom, no gravitational waves, no Newtonian interactions), only a finite number of topological ones. One important feature of 3-dimensional gravity comes from dimensional analysis (here we are considering the case $c=1$ for simplicity). In fact, it can be shown that the Newton constant $G_N$ has the dimension of inverse mass. This is easily seen by remembering that Newtonian gravity satisfies the local Gauss law 
\begin{align}
	\nabla \cdot \mathbf{g} = 4\pi G_N \rho
\end{align}
where $\mathbf{g}$ is the Newtonian acceleration. Assuming to have only two spatial dimension and considering for simplicity rotational symmetry, the above law implies 
\begin{align}
	|\mathbf{g}| = \frac{2 G_N m}{r}
\end{align}
which in turn means that 
\begin{align}
	\frac{L}{T^2}
	=
	[G_N] \frac{M}{L}
	\qquad
	\implies 
	\qquad
	[G_N] = \frac{1}{M}
\end{align}
where recall that the assumption $c=1$ also implies that $[c] = \frac{L}{T} = 1$. Therefore, the Newton constant $G_N$ has the dimensions of inverse mass, which means that classical gravity in 3 dimension already has a length scale built-in from the outset. As a consequence, keeping in mind Gauss motto, we expect that some kind of deformation has to be present already at the classical level. Skipping the details (which can for example be found in \cite{Arzano:2021scz} and in several literature papers) the way one proceeds is along the following steps.
\begin{itemize}
	\item[1)] The starting point is given by the action of a point particle coupled with gravity. As said above, the number of degrees of freedom of gravity in 3 dimension is finite, so we can actually get an explicit solution of the equations of motion  for them\footnote{This is not possible in general in 4 dimensions, which is the reason why this procedure cannot be repeated in the more realistic setting of a universe with $3+1$ dimensions.};
	\item[2)] Substitute these solutions back to the action of a point particle coupled with gravity, obtaining an effective action of a particle `dressed' in its own gravitational field;
	\item[3)] This action can then be written in a way analogous to the action in  \eqref{explicitdefaction}. 
\end{itemize}
The way to relate the above discussion about 3-dimensional gravity and quantum gravity is given by the following  argument \cite{Freidel:2003sp},\cite{Arzano:2021scz}. We will skip the technicalities, which can be found on the provided references, but the main idea of the argument can be summarized as follows.

Assume we have a spatially planar system in $3+1$ dimensions consisting of quantum particles coupled to the gravitational field. Then one can show that this system can be described as classical particles in $2+1$ dimensions. Hence they are described by 3-dimensional gravity, and (using the above procedure) the system will therefore be characterized by the deformation of spacetime symmetries. More precisely, the Hilbert space $\mathcal{H}^3$ of the system will have some symmetry group $\mathcal{P}^3$. At the same time, this original planar system is just a particular planar type of physical system in a $(3+1)$-dimensional spacetime. Therefore, the Hilbert space $\mathcal{H}^3$ of our subsystem is a subspace of $\mathcal{H}^4$, which is the vector field in the context of a full quantum theory of gravity. In particular, the symmetry group $\mathcal{P}^3$ must then be a subgroup of the full symmetry group $\mathcal{P}^4$ acting on $\mathcal{H}^4$. But now we know that the symmetries encoded in $\mathcal{P}^3$ are the deformed ones that we talked about, so that $\mathcal{P}^4$ cannot just be the Lorentz group (because in that case $\mathcal{P}^3$ couldn't be its subgroup). Therefore, we also expect that $\mathcal{P}^4$ (which we recall describes the symmetries of a full quantum gravity) must be some deformed symmetry group.

\section*{Acknowledgment}
This work was supported by funds provided by the Polish National Science Center,  the project number  2019/33/B/ST2/00050, and for JKG also by the project number 2017/27/B/ST2/01902.

\end{document}